\documentclass[twocolumn,aps,amsmath,amssymb,showpacs]{revtex4}

\usepackage{graphicx} 
\usepackage{epsfig}

\begin{document}

\title{Exact results for the Barab\'asi queuing model}
\author{C. Anteneodo} 
\affiliation{Departamento de F\'{\i}sica, PUC-Rio and  
National Institute of Science and Technology for Complex Systems,    
 Rua Marqu\^es de S\~ao Vicente 225, CEP 22453-900 RJ, Rio de Janeiro, Brazil }

%\date{\today}
\begin{abstract}

Previous works on  the queuing model introduced by  Barab\'asi to account for the 
heavy tailed distributions of the temporal patterns found in many human activities  
mainly concentrate on the extremal dynamics case and on lists of only two items. 
Here we obtain exact results for the general case with arbitrary values of the list 
length $L$ and of the degree of randomness  
that interpolates between the deterministic and purely random limits.
The statistically fundamental quantities are extracted from the solution of master equations. 
From this analysis, new scaling features of the model are uncovered.

\end{abstract}


\maketitle 

\section{Introduction}
 
Many human activities, such as mail and e-mail exchanges, library loans,    
stock market transactions~\cite{barabasi1}, or even motor activities~\cite{nakamura}, 
display heavy tailed inter-event and waiting time distributions.   
To account for these heavy tails~\cite{barabasi1}, a priority queuing model has 
been proposed by Barab\'asi~\cite{barabasi0}, 
that since then stimulated an active field of research with potential practical applications   
(e.g., see Refs.~\cite{aging,biaseddiff,exact2,cajueiro,transient}).

Within Barab\'asi priority queuing model (BQM), 
each item in a list of fixed length $L$ has a  priority value. 
At each time-step, the maximal priority task is executed with probability $p$, otherwise, 
a randomly selected one is accomplished.  Once a task is executed, it is substituted 
by a new one (or the same) that adopts a new randomly selected priority value drawn 
from a probability density function (PDF) $\rho(x)$. 
This simple model yields power-law tailed distributions of inter-events times, mimicking 
the empirical histograms of many human activities.

Besides the value of queuing models for diverse practical questions, 
another issue that makes BQM attractive is its connection with diverse other 
physical problems such as invasion percolation~\cite{transient,percolation} or 
self-organized evolutionary models~\cite{BS0,BS1,BS2}, 
as soon as the roles of priorities and fitness can be identified. 

However, exact results for the BQM, both for steady~\cite{exact2} and transient~\cite{transient} regimes,  
have been obtained for the simplest instance $L=2$ only.
Although lists of two items already display the power-law decay of the distribution   
of waiting times when $p$ approaches unity, naturally, other features are missed in the 
simplest case. 
Moreover, special attention has been given to the particular, and more tractable case, of  
extremal dynamics when $p\to1$~\cite{barabasi1,biaseddiff,percolation}, 
while non-null degree of randomness $(1-p\neq 0)$ may also 
display interesting features.
Then, in the present work, we tackle the BQM with arbitrary values of $p$ and $L$.

The manuscript is organized as follows. In the next section we show exact results for the 
PDFs of priorities in lists of arbitrary length $L$, by recourse to a master equation. 
In Sec. III we obtain an approximate expression for the 
waiting time distribution. Sec. IV deals with exact results for ``avalanches'' 
which provide the time that higher priority tasks (above a threshold)   
remain in the list, and is also related to waiting time duration. 
The last section contains final remarks.

\section{Exact treatment}

A fundamental quantity is  the probability 
that there are $n$ tasks with priority higher than a given value $x$, at time $t$, $P_{n,t}(x)$. 
Its time evolution is ruled by a master equation (ME) of the form  
\begin{equation}  \label{master}
P_{n,t+1}=M_{n,n+1}P_{n+1,t} + M_{n,n}P_{n,t}+M_{n,n-1}P_{n-1,t}\,, 
\end{equation}
for $n=0,1,\ldots, L$,  
with the non-null elements of the tridiagonal matrix $\mathbf{M}$ given by 
\begin{eqnarray} \nonumber
&&M_{n-1,n}(x) = px+(1-p) x n/L, \\ \nonumber
&&M_{n,n}( x) = p(1- x)+(1-p)\bigl( x(L-n)+(1- x)n \bigr)/L, \\ 
&&M_{n+1,n}( x) = (1-p)(1- x)(L-n)/L, \label{Mnn}
\end{eqnarray}
for $n=1,\ldots,L$,  and additionally  $M_{1,0}( x) = 1- x$, $M_{0,0}( x)=  x$.  
Here we have taken $\rho(x)=1$, however  generality can be 
recovered simply by redefining the threshold through $x\to  R(x)=\int_0^x\rho(x')dx'$.

Notice that the ME (\ref{master})-(\ref{Mnn}) signals a biased random walk with reflecting 
boundaries at $n=0$ and $n=L$, setting the basis to write a continuum limit approximation. 
However, for arbitrary $L$, 
drift and diffusion coefficients are state dependent  and  the approach of 
biased diffusion successfully applied~\cite{biaseddiff} to determine the scaling of the 
waiting time distribution, in other queuing systems with constant coefficients, 
becomes more tricky in the non-deterministic case  $p\neq 1$.

Then, let us find the exact steady solution of the ME  
(\ref{master})-(\ref{Mnn}) for arbitrary length $L$. By recursion, one gets  
\begin{equation} \label{Pfinite} 
P_n( x) = \frac{L!\Gamma(a+1)(1- x)^n}{(L-n)!\Gamma(a+n+1)(1-p) x^n}P_0( x),
\end{equation}  
for $1\le n\le L$, where $a=pL/(1-p)$, and  from normalization     
\begin{equation}   \label{P0}
P_0( x) = \bigl( 1+ \sum_{n=1}^L P_n( x)/P_0( x)  \bigr)^{-1} \,.
\end{equation}

The distribution $P_n$, given by Eqs.~(\ref{Pfinite})-(\ref{P0}), 
can be used now to evaluate diverse 
meaningful quantities. In particular, the PDF of the $n$th largest priority value can be 
extracted from the condition $\int_0^ x p_n( x')d x'=\sum_{m=0}^{n-1}P_m( x)$, hence   
\begin{equation} \label{Porder}
p_n( x)=\frac{\partial}{\partial  x} \sum_{m=0}^{n-1}P_m( x)\,.
\end{equation}

\begin{figure}[h!]
\centering 
\includegraphics*[bb=150 270 520 650, width=0.5\textwidth]{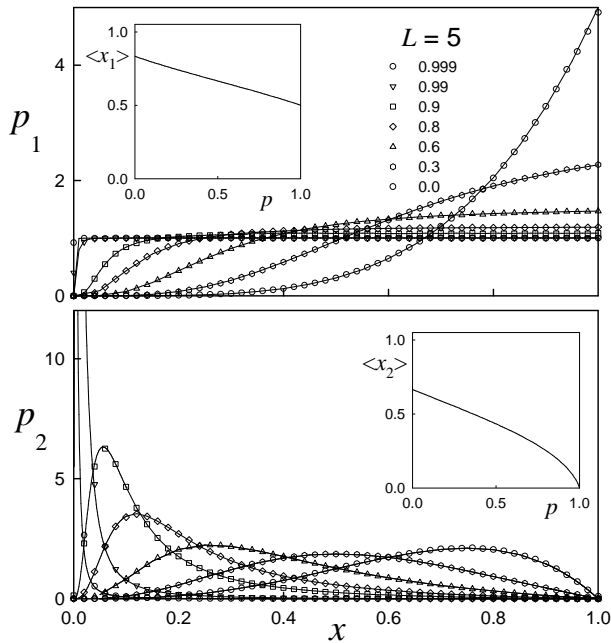}  
 \caption{PDFs of the largest (upper panel) and the second largest (lower panel) 
 priority values,  for $L=5$, different values of  $p$  indicated on the figure and $R(x)=x$. 
 Solid lines correspond to exact results and symbols  to numerical simulations of the BQM  
 performed as in previous figures.
 In the insets, the average values are displayed as a function of $p$.
  }
\label{fig:first5}
\end{figure}

Fig.~\ref{fig:first5} shows the exact PDFs of the two largest priorities in the list, 
$p_1(x)=P^\prime_0(x)$ and $p_2(x)=P^\prime_0(x)+P^\prime_1(x)$, for  
$L=5$ and different values of $p$, compared to the results of 
numerical simulations of the BQM.

In the fully random case $p=0$, Eqs.~(\ref{Pfinite})-(\ref{P0}) yield 
$P_n( x)=\binom{L}{n}(1- x)^{n} x^{L-n}$,
hence $p_n( x)=L\binom{L-1}{n-1}(1- x)^{n-1} x^{L-n}$, in accord with  
straightforward combinatorial analysis. 
In the opposite limit $p\to 1$, $p_1( x)$ gets closer to a unit step function at $x=0$ while 
$p_2( x)$ approaches the Dirac delta function $\delta(x)$. 
This is expected since those tasks that have entered the list more recently and adopted 
priority values uniformly distributed in [0,1]  
have more chances to be chosen again, while the older tasks are more and more 
likely to remain in the list forever as $p$ tends to $1$, then 
the second priority value (and together with it the remaining ones)  
collapse to zero.

\begin{figure}[hb!]
\centering 
\includegraphics*[bb=110 240 530 670, width=0.5\textwidth]{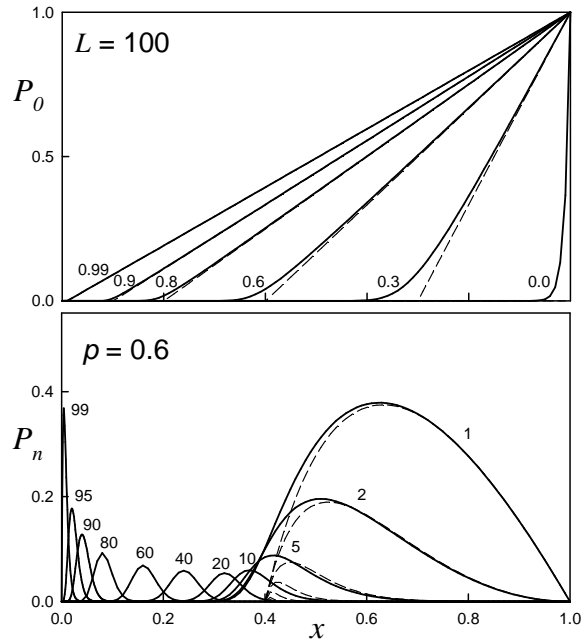}  
 \caption{Probabilities $P_0(x)$ (for different values of $p$, upper panel) and 
 $P_n(x)$ (for different values of $n$ and $p=0.6$, lower panel), at $L=100$ and $R(x)=x$.  
 Solid lines correspond to exact results, dashed lines to the large $L/(1-p)$ approximation. 
  }
\label{fig:P0}
\end{figure}

For large enough $L$ (namely, $L/(1-p)>>1$), Eqs.~(\ref{Pfinite})-(\ref{P0}) 
lead to $p_1 \simeq H(x-1+p)/p$, 
where $H$ is the Heaviside unit step function, and 
$p_2\simeq (1-p)(1/x^2-1)H(x-1+p)/p^2$. 
In fact, finding directly 
the steady state solution of the ME (\ref{master})-(\ref{Mnn}),  
in the limit of large $L$ for fixed $n$ (hence neglecting terms of order $n/L$), 
or also when $p\to 1$,  
one obtains  a geometric progression 
that, for $ x>1-p$, can be summed up to obtain the simple expression
\begin{equation} \label{P0infinite} 
P_0(x) \simeq ( x-1+p)/p \;\;\;\; \mbox{and} 
\end{equation}
\begin{equation}  \label{Pinfinite}
P_n( x) \simeq \frac{( x-1+p)(1-p)^{n-1}(1- x)^{n}}{p^{n+1} x^n},\;\;\; \mbox{for $0<n\le L$} . 
\end{equation} 
For $x\le 1-p$, all $P_n$ tend to vanish in the large $L$ limit.  
Fig.~\ref{fig:P0} illustrates the performance of this approximation in comparison with exact results. 
The assumption $n<<L$ fails as soon as the probability 
that $n>{\cal O}(1)$ becomes non negligible. 
For each $x<1-p$, the exact $P_n$ is peaked around  $n\simeq (x-1+p)L/(1-p)$.  
The approximation becomes exact both in the limits of $L\to \infty$ and $p\to 1$.

The PDF of all priorities $x$ in the list, $p(x)$ 
verifies $\sum_{n=1}^{L}p_n(x)=Lp(x)$. Its  time evolution is given by
\begin{equation} \label{pt}
p(x,t+1)=p(x,t)+\bigl(\rho(x)-p p_1(x,t)-(1-p) p(x,t) \bigr)/L, 
\end{equation}
that in the long-time limit leads to the relation
\begin{equation} \label{Pgen}
pP_0(x)+(1-p)P(x)=R(x),
\end{equation}
where $P(x)=\int_0^x p(x')dx'$.

Let us call old tasks those items whose priority has not been assigned at a given step.    
The cumulative PDF of old task  priorities, $O(x)$, can be obtained from 
the relation 
\begin{equation}
L P(x)=R(x)+(L-1)O(x)
\end{equation}
and,  by means of Eq.~(\ref{Pgen}) can be expressed as  
\begin{equation}  \label{allold}
O(x) =\frac{(L+p-1)R(x)-pLP_0(x)}{(L-1)(1-p)}\,. 
\end{equation}
 
In the particular case $L=2$, Eqs.~(\ref{Pfinite})-(\ref{P0}) give 
$P_0( x)=(1+p) x^2/(1-p+2p x)$ and recalling that its derivation  
was carried out for uniform $\rho(x)$ but the general case is  recovered 
simply through the mapping $ x \to R(x)$, then, Eq.~(\ref{allold}) allows to re-obtain the result of 
Vazquez~\cite{exact2}, namely, $O(x)= (1+p)R(x)/[1-p+2pR(x)]$.

\begin{figure}[h!]
\centering 
\includegraphics*[bb=150 360 530 620, width=0.5\textwidth]{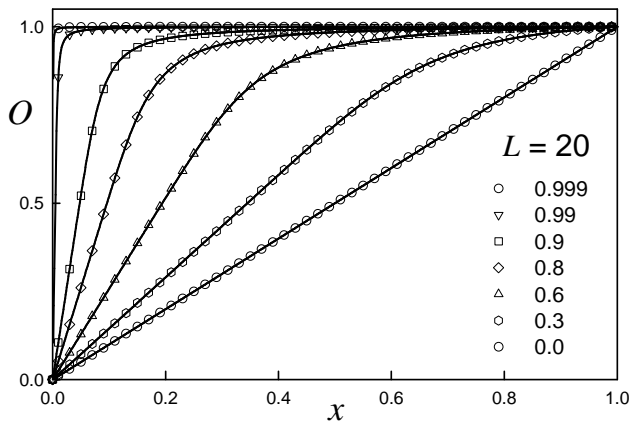}  
 \caption{Cumulative PDFs of old task priorities, 
 for $L=20$, different values of  $p$ and $R(x)=x$. 
Symbols correspond to numerical simulations of the BQM performed as in previous figures,  
black lines to the exact results from Eq.~(\ref{allold}). 
 }
\label{fig:old20}
\end{figure}

Fig.~\ref{fig:old20} illustrates the behavior of $O(x)$ for different values of $p$ and $L=20$. 
The distribution of the bulk of old values for arbitrary $L$ is qualitatively similar to 
that obtained for $L=2$ in Ref.~\cite{exact2}. 

For $L>2$, however, $P(x)$ and $O(x)$ are mean-field quantities, while   
more meaningful is the distribution of the largest old priority 
$O_1(x)$ (that, of course, for  $L=2$ coincides with $O(x)$). 
It verifies 
\begin{equation}  \label{O1}
P_0(x)=R(x)O_1(x),
\end{equation}
since the probability that there are no tasks above $x$,  $P_0(x)$, 
is the product of  the probability that the freshly assigned (new) priority value 
is below $x$ times the probability 
that the highest old task priority (hence also the remaining ones) is below $x$, 
as soon as the new priority value and the old ones are independent. 
For the particular case $p=0$, $O_1(x)=R^{L-1}(x)$ 
%and $O_2(x)=[L-1-(L-2)R(x)]R^{L-2}(x)$,
while in the opposite limit $p\to1$, it  tends to a unit step function at $x=0$.

For $L>2$, the distribution of the second largest old priority $O_2(x)$ 
can be extracted from the identity 
\begin{equation} \label{O2}
P_1=(1-R) O_1 +R(O_2-O_1), 
\end{equation}
which comes from considering 
that the probability that there is only one task above $x$, $P_1(x)$, is 
prob.[new $\ge x$ $\wedge$ 1st old $\le x$]) + prob.([new $\le x$  
$\wedge$ 2nd old $\le x$ $\wedge$ 1st old $\ge x$]), while  
prob.(2nd old $\le x$  $\wedge$ 1st old $\ge x$)=prob.(2nd old $\le x$)-
prob.(2nd old $\le x$ $\wedge$ 1st old $\le x$)=$O_2(x)-O_1(x)$, 
since the first and second largest old values are not independent. 
Analogously, in general one has 
\begin{equation} \label{On}
P_n=(1-R)(O_n-O_{n-1})+R(O_{n+1}-O_n), 
\end{equation}
for $1\le n\le L-1$, taking $O_0=0$ and $O_L=1$, while $P_L=(1-R)(1-O_{L-1})$.
From where the whole family of stationary old task distributions can be 
straightforwardly obtained. 

\begin{figure}[h!]
\centering 
\includegraphics*[bb=140 200 530 620, width=0.5\textwidth]{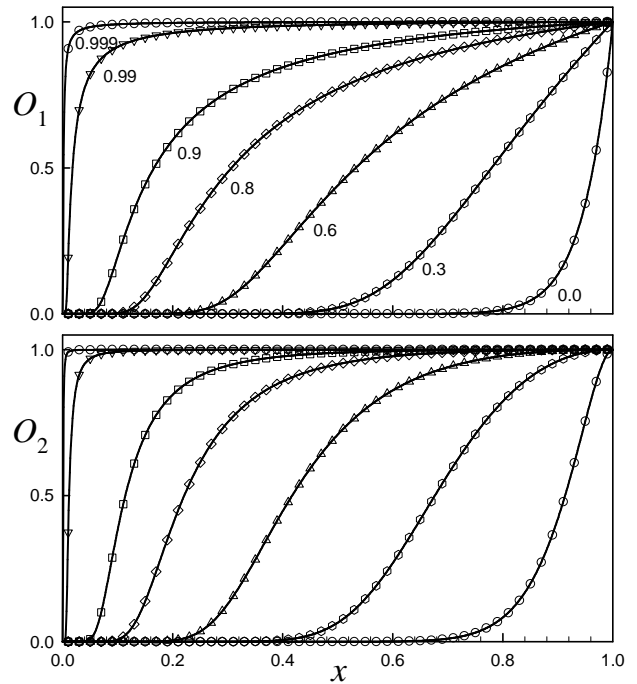}  
 \caption{ Cumulative PDFs of the first (upper panel) and second 
 (lower panel) largest old task priorities, 
 for $L=20$, different values of  $p$ and $R(x)=x$. 
Symbols correspond to numerical simulations of the BQM performed as in previous figures,  
black lines to the exact results from Eqs.~(\ref{O1})-(\ref{O2}). 
 }
\label{fig:all20}
\end{figure}

Exact results for $O_1(x)$ and $O_2(x)$ are compared to the outcomes of numerical simulations of the BQM 
in Fig.~\ref{fig:all20}, for $L=20$, different values of $p$ and $R(x)=x$. 
Observe that $O_2(x)$ is bounded 
from below by $O_1(x)$ and more generally $O_1(x)\le O_2(x)\le \ldots \le O_{L-1}(x)$.

\section{Waiting time distribution}

The family of distributions of old values $\{O_n, 1\le n\le L-1)\}$  
should  in principle allow to compute exactly the distribution of waiting times $P_w(\tau)$ 
in the steady regime. 

$P_w(\tau)$ can be obtained as
\begin{equation} \label{Ptau}     
P_w(\tau)= \int_0^1 dR(x) r_\tau(x),  
\end{equation}
where $r_\tau(x)$ is the probability that a task (let us call it $X$) with freshly 
acquired priority value $x$ at a given 
time $t=t_o$ (once attained the steady state) is again selected for the first time at $t=t_0+\tau$.

For $\tau=1$, 
\begin{equation} \label{rtauone}     
r_1(x)= p O_1(x)+(1-p)/L\,.  
\end{equation}
Hence $P_w(1)=1/L$ when $p=0$ and it tends to one in the opposite case $p\to 1$.
By means of the approximate Eq.~(\ref{P0infinite}) for $P_0(x)$, one has 
$P_w(1)\simeq p+(1-p)\ln(1-p)+(1-p)/L$.

The probability that, instead of $X$, the first old task is selected  at $t_0+1$   
is $p(1-O_1(x))+(1-p)/L$, while the probability that any other old 
task is selected (for $L>2$) is $(1-p)/L$.   
Given each of these $L-1$ cases, for computing $r_2(x)$, one has in principle a different probability 
of selection of $X$ at the second step ($t=t_0+2$) which will be a function of $O_1$ and $O_2$.
More generally, the exact calculation of $P_w(\tau)$, for $\tau>1$, will require to consider 
a branching process, with $L-1$ paths at each node, such that for $L>2$, $r_\tau(x)$  does not factorize. 
This tree generalizes the branching process considered in analogy to invasion percolation 
for $L=2$~\cite{transient}. 

In the first steps (up to $\tau\sim L$), the statistics will be conditioned by the memory of 
previous selections (aging regime). This is because recently chosen tasks have propensity 
(the higher, the closer $p$ to 1) to be 
chosen again, dominating $P_w$ at small $\tau$.

\begin{figure}[ht!]
\centering 
\includegraphics*[bb=100 420 530 670, width=0.5\textwidth]{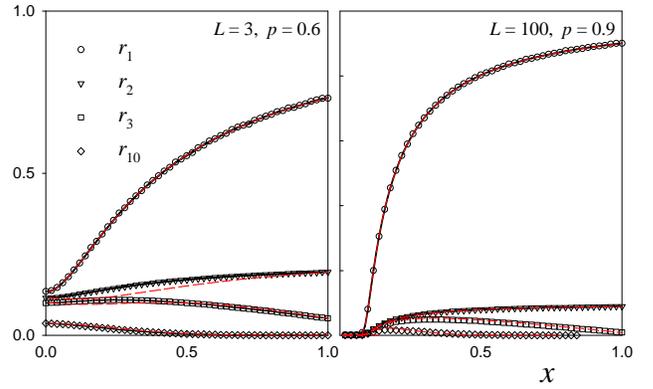}  
 \caption{(Color online) Integrands $r_\tau$ of the distribution of waiting times, for values 
 of $\tau$ indicated on the figure and two couples of parameters $(L,p)$.
 Symbols correspond to numerical simulations, solid lines to exact results and 
 dashed red lines to the approximate analytical expressions.  
 }
\label{fig:f_rrr}
\end{figure}

\begin{figure}[hb!]
\centering 
\includegraphics*[bb=130 310 510 680, width=0.5\textwidth]{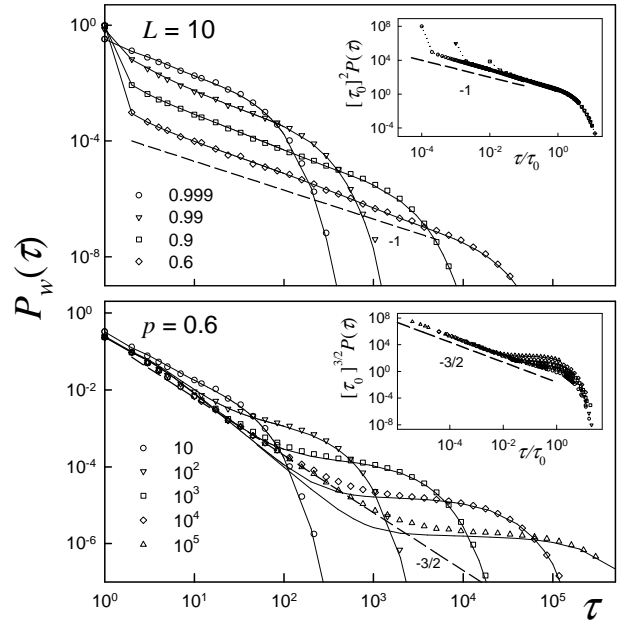}  
 \caption{Distributions of waiting times, for fixed size $L=10$ and different  values of $p$ indicated 
 on the figure  (upper panel) and for fixed $p=0.6$ and increasing values of $L$ indicated on 
 the figure (lower panel), $R(x)=x$. Solid lines join the analytical results from Eqs.~(\ref{Ptau}) 
 and (\ref{rtau}) and 
 symbols correspond to numerical simulations of the BQM. 
 Insets: rescaled plots of the numerical histograms, where $\tau_0=1/\ln(L/(L-1+p))$.  
 Dashed lines with slopes  -1 (upper inset) and  -3/2 (lower inset) are drawn for comparison.  
 }
\label{fig:f_tau}
\end{figure}

In particular, for $\tau=2$ one obtains  
\begin{equation} \label{rtautwo}     
r_2(x)=c(1-r_1)+pR[(p+c)O_2+\{c(L-2)-p\}O_1] \,,  
\end{equation}
where $c=(1-p)/L$. 
For the special case $L=2$, $O_2=1$, then  one recovers 
the expression   found in Ref.~\cite{exact2} for $L=2$, namely, $r_2=(1-r_1)[pR+(1-p)/2]$.

At further time-steps this is a hard trail to proceed and the results may not be expressible in 
a readily manageable form. 
However, notice that while for small $\tau$, the integral of $r_k$ is 
dominated by large values of $x$, due 
to the propensity of such values to be re-chosen early, contrarily, 
for large enough $\tau$, $r_\tau$ (and hence $P_w$) 
will gain the main contribution from  the purely random (unconditioned) selection from the bulk of 
relatively small $x$ values (as can be seen in Fig.~\ref{fig:f_rrr} where $r_\tau(x)$ is 
displayed). This is expected to apply also when $p\to 1$ at any $\tau$. 
For such cases,  one can write
\begin{equation} \label{rtau}     
r_\tau(x)\simeq   (1-r_1(x))(1-f(x))^{\tau-2}f(x),  
\end{equation}
where  $f(x)$ is the effective probability that task $X$  
 is selected at some given step $t>t_0+1$ and can be estimated as 
 $f(x)=p P_0(x) +(1-p)/L$, as soon as $P_0=RO_1$ is the probability that 
 there are no tasks with priorities higher than $x$. 
Fig.~\ref{fig:f_rrr} also exhibits the comparison between exact and approximated 
functions $r_\tau$.

In particular, for $p=0$, Eq.~(\ref{Ptau}) is independent of the choice of $f(x)$ and it  
correctly  yields the pure exponential decay $P_w(\tau)= (1-1/L)^{\tau-1}/L$ for 
all $\tau$~\cite{barabasi0}. 
In the opposite case $p\to 1$, and using the approximation 
given by Eq.~(\ref{P0infinite}) for $P_0$,  Eq.~(\ref{Ptau}) leads to the asymptotic behavior   
\begin{equation}
P_w(\tau) \sim \frac{1}{\tau}\exp(-\tau/\tau_0),
\end{equation}
where $\tau_0=1/\ln(L/(L-1+p))\sim L/(1-p)$. 

This expression for the characteristic time $\tau_0$ applies por any $p$. 
Thus, the characteristic  exponential decay time $\tau_0$ 
is shifted to larger $\tau$ when $p\to 1$ as well as when  $L$ increases.

Analytical predictions are 
compared to numerical simulations in Fig.~\ref{fig:f_tau}. One observes that the 
approximate expression derived from Eq.~(\ref{Ptau}) manages to 
describe the exponential cutoff in all cases and the scaling regime  in 
the limit $p\to 1$, although it fails to predict the   -3/2 power-law   neatly observed in 
numerical simulations for $0<p<1$ as $L\to \infty$ (notice in the lower panel of Fig.~\ref{fig:f_tau} 
the deviation for $\tau\lesssim L$,  leading to a spurious power-law exponent -2). 
This is due to the fact that the aging regime  is overlooked by this approximation. 
Let us remark that a -3/2 exponent is also found in classical queuing models with 
fluctuating length~\cite{biaseddiff,barabasi0} 
and the return time distribution of a random walk is at its origin.  
In view of the difficulties to find the exact expression for $P_w(\tau)$, to explain 
this scaling regime, we will solve next a closely 
related problem.

\section{Avalanches}

Let as also consider now the events between  two successive times when the number $n$ of 
priorities above 
a given threshold $ x$ vanishes (avalanche). Avalanche duration is relevant in the present 
context as  as soon as it provides the duration of intervals in which  
there are queued tasks with priorities above a threshold to be executed.  From the 
viewpoint of random walks, this is a first passage problem.
Following the lines in Ref.~\cite{BS2}, let us define $Q_{n,t}( x)$, 
the probability of having $n$ values with priorities higher than $ x$, 
given that an avalanche started at $t=0$  ($t$ time units ago). 
$Q_{n,t}$ follows the same ME (\ref{master})-(\ref{Mnn}) as $P_{n,t}$ does, except for $M_{0,1}=0$, 
and the initial condition is $Q_{1,0}=1- x$ and $Q_{n,0}=0,\;\forall n>1$.    
Thus, the probability that an avalanche, relative to threshold $x$, has duration $t$ is 
\begin{equation} \label{avalanche}
q_t( x)=   
x Q_{1,t-1}( x).
\end{equation}

\begin{figure}[h!]
\centering 
\includegraphics*[bb=110 420 550 700, width=0.5\textwidth]{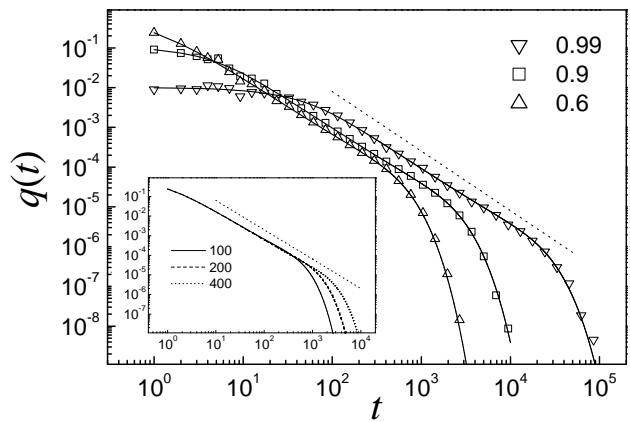}  
 \caption{Distribution of avalanche size   $q(t)\equiv q_t(x=1-p)$
 for $L=100$, different values of  $p$ indicated on the figure and $R(x)=x$. 
 Solid lines join the exact values  and symbols correspond to numerical simulations of the BQM.  
In the inset,  exact results for $p=0.6$ and different values of $L$ indicated 
on the figure are displayed. 
Dotted straight lines with slope -3/2 are drawn for comparison. 
 }
\label{fig:avalanche}
\end{figure}

Fig.~\ref{fig:avalanche} illustrates the scaling that comes up for any $p$ at the critical 
threshold $ x=1-p$. 
Exact results were obtained by numerical integration of the ME for $Q_n$ and compared 
to the results of numerical simulations of the BQM. Notice that the scaling region increases 
with $L$ and shifts towards larger times as $p\to 1$.

The ME of $Q_n$ can be solved analytically through diverse standard 
methods~\cite{feller,redner}. 
Yet, in the limit of large $L$ and fixed $n$, the ME describes a simple biased random walk, with 
an absorbing boundary at $n=0$ and probabilities to step  either to the right, 
to the left, or remain still, given by  $m^+=(1-p)(1-x), \;m^-=px, \;m^0=1-m^+-m^-$,  respectively. 
From this viewpoint, $q_t(x)$ is the probability that the 
first return to the origin occurs at $t$ when the avalanche started at $t=0$,  
while $Q_{1,t}(x)$ is the probability of reaching $n=1$ at time $t-1$, without having visited $n\le 0$. 
Thus, $q$ just differs from $Q_1$ in appending the last step from 1 to 0.
For any $n$, $Q_{n,t}(x)$ can be found by solving first the unbounded problem  and then 
resorting to the reflection principle~\cite{fisher,khantha}.
Moreover, if we are concerned with the asymptotic behavior, we can directly take advantage of 
the Gaussian approximation from the central limit theorem. Therefore, one has 

\begin{equation}
Q_{n,t}(x)\simeq (1-x)\frac{e^{-\frac{(n-1-c t)^2}{2\sigma^2 t}}-\frac{m^-}{m^+}
e^{-\frac{(n+1-c t)^2}{2\sigma^2 t} }}{\sqrt{2\pi \sigma^2t}} \,,
\end{equation}
 where $c$, and $\sigma^2$ are the mean and variance of each single step. 
This readily leads to the asymptotic behaviors
\begin{equation}
q_t(x)\sim  \left\{ \begin{array}{ll}
              t^{-3/2}, &\mbox{if $c=0$}\cr
             t^{-1/2}\exp(\frac{-c^2 t}{2\sigma^2}), &\mbox{otherwise},
           \end{array} \right. \\
\end{equation}
that is, an exponential decay dominates the long-time decay in the biased cases, 
meanwhile, if $c\equiv m^+-m^-=0$ (hence $x=1-p$), a power-law arises in the large $L$ limit, 
in agreement with the results displayed in Fig.~\ref{fig:avalanche} and with the 
well known results for a driftless random walk~\cite{fisher}. 
In particular, there is a correspondence with the random annealed Bak-Sneppen model, where the same 
scaling is observed for any $K$ at the critical threshold $ x=1/K$~\cite{BS2}. 
Let us remark that in the Bak-Sneppen model the transition matrix for 
the associated ME has $2K$ non-null diagonals, 
and a generic  univoque  relation  between $p$ and $K$ does not emerge.
However, concerning avalanches,  
the equivalence between both models arises for $K=1/(1-p)$.  
Due to the threshold being  an upper or lower bound in each case,  
that relation is complementary to $K=1/p$ which arises   
by identifying ratios of deterministic/random sites~\cite{barabasi1}.

\section{Final comments}

Summarizing, we obtained analytical results for the BQM with queues of arbitrary length. 
Exact expressions were shown to be in  agreement with the 
outcomes of numerical simulations of the dynamics.  
Progress has still to be made to obtain the exact waiting time distribution 
that displays different regimes    
between the purely exponential one (at $p=0$) and the power-law decay with unit exponent 
(at $p\to1$), when $L\to \infty$. However, an approximate expression has been found that 
accounts for most of the distribution traits.  
Moreover, we have shown that avalanches, 
at the critical threshold $ x=1-p$,  constitute another scale-free feature of the BQM for $p>0$.
Besides the main applications here illustrated, the present results may allow 
to estimate many other relevant statistical quantities of the BQM and can be 
extended to other queuing systems. 
Furthermore, our exact results set the basis to further explore the correspondence 
between BQM and  other related models.

{\bf Acknowledgements:}   
C.A. acknowledges D.R. Chialvo and R.O. Vallejos for useful suggestions and discussions, 
and  Brazilian agencies CNPq and Faperj for partial financial support.

\end{document}